\documentclass[11pt]{article}
\usepackage{authblk}
\usepackage{amsmath,amssymb,dsfont}
\usepackage{hyperref}
\usepackage[utf8]{inputenc}
\usepackage{slashed}

\newcommand{\half}{\frac{1}{2}}
\newcommand{\JJ}{\mathds{J}}
\newcommand{\KK}{\mathds{K}}
\newcommand{\DD}{\mathds{D}}
\newcommand{\TT}{\mathds{T}}
\newcommand{\ZZ}{\mathds{Z}}

\newcommand{\QQ}{\mathds{Q}}
\newcommand{\QQb}{\overline{\mathds{Q}}}
\newcommand{\AAA}{\mathds{A}}
\newcommand{\FF}{\mathds{F}}
\newcommand{\calF}{\mathcal{F}}
\newcommand{\calG}{\mathcal{G}}
\newcommand{\calH}{\mathcal{H}}
\newcommand{\calL}{\mathcal{L}}

\newcommand{\se}{\slashed{e}}
\newcommand{\sT}{\slashed{T}}
\newcommand{\sD}{\slashed{D}}

\newcommand{\lD}{\overleftarrow{D}}

\newcommand{\psibar}{{\overline{\psi}}}

\newcommand{\gf}{\gamma_5}
\newcommand{\dV}{|e|d^4x}

\newcommand{\MP}{{M_\text{P}}}
\newcommand{\gaugeG}{$SO(3,1)\times SU(2)\times U(1)$ }

\newcommand{\cecs}{Centro de Estudios Científicos (CECs) Av.~Arturo Prat~514, Valdivia, Chile}
\newcommand{\uantof}{Departamento de Física, Universidad de Antofagasta, Aptdo. 02800, Chile}

\numberwithin{equation}{section}
\begin{document}

\title{Chiral gauge theory and gravity from unconventional supersymmetry}

\author[1]{Pedro D. Alvarez\thanks{pedro.alvarez@uantof.cl}}
\affil[1]{\uantof}

\author[2]{Mauricio Valenzuela\thanks{supercuerda@gmail.com}}
\affil[2]{\cecs}

\author[2]{Jorge Zanelli\thanks{z@cecs.cl}}


\maketitle

\begin{abstract}

From a gauge $SU(2,2|2)$ model with broken supersymmetry, we construct an action for $SU(2)\times U(1)$ Yang-Mills theory coupled to gravity and matter. The connection components for AdS boosts and special conformal translations are auxiliary fields and their fixing reduces the theory to two distintive sectors: a vector-like gauge theory with general relativity and a chiral gauge theory where gravity drops out. We discuss  some of the main classical features of the model such as the predicted tree level gauge couplings, cosmological constant value, mass-like terms and the Einstein equations.
\end{abstract}
\pagebreak

\tableofcontents

\section{Introduction}        

Supersymmetry (SUSY) is the largest symmetry of Quantum Field Theory that unifies spacetime and internal symmetries in a nontrivial manner \cite{HLS, Kac}, circumventing the no-go theorems of Coleman and Mandula \cite{Coleman-Mandula}. SUSY generically predicts the necessary presence of fermions and bosons, improves renormalizability, produce viable dark matter candidate models and, if promoted to a local symmetry, can include gravity. For historical and technical reviews, see for example, \cite{Weinberg,Nilles:1983ge,deWit:2002vz}. 

The realization that softly broken SUSY can stabilize the mass of the Higgs boson \cite{Gildener:1976ai,Susskind:1978ms}, along with gauge coupling unification \cite{Amaldi:1991cn,Ellis:1990wk,Langacker:1991an}, provides a motivation for SUSY unified extensions of the Standard Model \cite{Witten:1981nf,Dimopoulos:1981zb,Sakai:1981gr,Kaul:1981wp}. The minimal phenomenologically viable model, the Minimal Supersymmetric Standard Model \cite{Dimopoulos:1981zb,Sakai:1981gr}, duplicates the particle content of the Standard Model and predicts the masses of the SUSY partners not far above the weak scale when considered as a solution to the hierarchy problem of the scalar sector of the Standard Model \cite{Barbieri:1987fn}. However, `natural' supersymmetric models that predict superpartners at the TeV scale are challenged by the lack of conclusive evidence for direct production in the Large Hadron Collider (LHC) so far \cite{Sirunyan:2019zfh,Uno:2019vxd}.

Even simplified models of SUSY have a large number of free parameters, $\sim \mathcal{O}(10)-\mathcal{O}(100)$. It is therefore hard to constrain their parameter space, although by considering more intricate models or relaxing some assumptions, the appearance of super partners can be postponed to higher energies \cite{Papucci:2011wy,Baer:2015rja}.

The possibility of not finding superpartners has motivated the proposal of non-simplified SUSY models that could still be viable solutions to the hierarchy problem\footnote{Other alternatives to SUSY in dealing with the hierarchy problem induced by a fundamental scalar are technicolor \cite{Susskind:1978ms,Dimopoulos:1979es,Weinberg:1975gm}, extra-dimensions \cite{Randall:1999ee} or, more recently, the idea of the relaxion \cite{Graham:2015cka}.} \cite{Baer:2015rja,Martin}. At the root of the problem is the fact that the underlying theory for SUSY breaking is not known \cite{Weinberg,Witten:1981nf,OReifeartaigh,Witten:1982df}. This has recently led to consider the possibility that unknown correlations among the parameters in the underlying high energy scale SUSY theory may be behind the automatic cancellations that would invalidate the assumption that superpartners should be light \cite{Tata}. 

There are certain features that are common to a large class of SUSY models. One of these features is the mass degeneracy in super Poincar\'e models, which is implied from the commutator between the momentum and the supercharge, $[P_\mu, Q^{\alpha}] =0$. Another common feature is that SUSY and internal gauge transformations also commute. This assumption is justified because matter and interaction fields in the Standard Model are in different representations of the gauge group and they should therefore belong to different supermultiplets. As a result, in conventional SUSY models all particle states are duplicated: for each particle there must be a SUSY partner with the same mass and gauge charges, but differing by half a unit of spin.

Following \cite{AVZ,APZ}, here we advocate a way of combining gauge bosons and matter fermions in a (super) Lie-algebra valued gauge connection that allow us to circumvent some of the above assumptions. This unconventional form of SUSY (uSUSY) still retains the spirit of a gauge symmetry principle in the sense that it unites spacetime and internal symmetries. As a consequence, this allows the construction low energy models with very few free parameters.

Having gauge bosons and matter fields transforming in the same multiplet imply that the SUSY generators have to be charged with respect to the gauge symmetry, hence their anti-commutators generate both spacetime and internal transformations,
\begin{equation} \label{u-susy}
\{ \overline{Q} , Q \} \sim \left(\;\JJ_{AdS} \;\;   \right) + \left(\; \TT_{Internal} \;  \right).
\end{equation}
Now, since the supercharges do not commute trivially with $AdS$-transvections, the different elements in the multiplet do not need to have the same masses.

Supergravity models can be constructed using superalgebra-valued gauge connections, as in the pioneer work of Macdowell and Mansouri \cite{MacDowell:1977jt} an attempt to construct a gauge theory for $OSp(4|1)$. Similar superalgebras have also been extensively explored in the context of odd-dimensional supergravity (SUGRA), where the gravitino enters naturally as part of the connection for the superalgebra \cite{TZ1,TZ2,HTZ,HZ}. This leads to a completely different family of locally supersymmetric theories that extend gravitation without SUSY partners for each field and no matching degrees of freedom between bosons and fermions.

The minimal SUGRA action that includes the Einstein-Hilbert and Rarita-Schwinger terms, enjoys the local supersymmetry \cite{Freedman-van Proeyen},
\begin{equation}\label{susytran}
\delta\chi_\mu= \partial_\mu\epsilon\,,\qquad  \delta e^a_\mu=\bar{\chi}_\mu\gamma^a \epsilon,
\end{equation}
where $\chi$ is the gravitino field, $e^a_\mu$ is the vielbein and $\epsilon$ is the fermionic parameter of the transformation. The gravitino belongs to the $1 \otimes 1/2 = 3/2 \oplus 1/2$ reducible representation of the Lorentz algebra, when not subsidiary conditions are imposed. The spin-$1/2$ field is often gauged away by a suitable choice of the parameter $\epsilon$ in \eqref{susytran}. It follows that the dynamical content of the spin-$1/2$ piece of the Rarita-Schwinger field should be trivial. Let us consider, however, the scenario in which supersymmetry is broken at low energies. This will prevent us to gauge away the spin-$1/2$ sector of the gravitino using \eqref{susytran} and therefore it may acquire non-trivial dynamics.

These statements motivated the search of a concrete mechanism by means of which the spin-$1/2$ sector of the gravitino may emerge as an observable particle in a broken phase of supergravity  \cite{AVZ,APZ,APRSZ,Gomes-Helayel,ACDAT,Andrianopoli:2019sip,ACGT}. These theories implement the ``matter ansatz'' for the gravitino, i.e. 
\begin{equation} \label{projection}
\chi_\mu= \gamma_\mu \psi, \qquad \psi=\frac{1}{D} \gamma^\mu\chi_\mu,
\end{equation}
where $\psi$ is a spinor zero-form containing the spin-$1/2$ projection of the gravitino and $D$ is the spacetime dimension. This ansatz allows to accommodate matter into a gauge connection.

As shown in \cite{GPZ}, starting with a three dimensional Chern-Simons theory for a superconnection --with zero propagating degrees of freedom--, the ansatz \eqref{projection} yields a propagating spin-1/2 field. This example shows that it is indeed possible for a dynamical spin-$1/2$ field to emerge as a Dirac particle from the part of the gravitino that is usually projected out. 

Here we focus on constructing a four dimensional N=2 uSUSY model equipped with a mechanism to obtain an $SU(2)\times U(1)$ chiral gauge theory, a minimal requirement to construct phenomenologically viable particle models. The model also contains another sector with a vector-like gauge theory and gravity.

The paper is organized as follows. In section \ref{model}, we describe the fundamentals of using USUSY to construct a model invariant under \gaugeG. In section \ref{physicalcontent}, we show how the system behaves around a dS or AdS vacua and elaborate on the physical contents of the theory. In section \ref{summary}, we summarized our results.

\section{The model}\label{model}

We will define a gauge theory that includes gravity and a gauge symmetry that can accommodate the electroweak sector by means of the superextension of a bosonic symmetry algebra that contains $SO(3,1)\times SU(2)\times U(1)$. The smallest superalgebra that fulfills this requirement is $su(2,2|2)$ and we consider the explicit representation given in Appendix A. The model includes matter in the form of a fermionic field in the fundamental representation of the bosonic gauge group. The novel feature of uSUSY, these matter fields $\psi_i$ are directly included in the gauge connection $\AAA$ for the super extended algebra,
\begin{equation}\label{Afermion}
\AAA_\psi:=\QQb^i \se \psi_i+\overline{\psi}^i \se \QQ_i \ \subset \AAA\ \,,
\end{equation}
where the spin 1/2 fermion 0-form $\psi_i$ is combined with the tetrad 1-form $\se := \gamma_a e^a= \gamma_a e^a_\mu dx^\mu$ (here $a$ is a flat spacetime index of the tangent space), and $\gamma_a$ are the Dirac matrices.\footnote{Here $\{\gamma^a,\gamma^b\}=2 \eta^{ab}$, where the flat spacetime metric is $\eta=\mathrm{diag}(-,+,+,+)$. The spinor indexes will be often omitted. The Dirac adjoint is defined by $\psibar_i=i\psi_i^\dag  \gamma^0$.} The SUSY generators $\QQ$ and $\QQb$ carry spinors (Greek) indices and belong to the fundamental representation of the internal symmetry group (Latin indices). This is to accommodate $\psi_i^\alpha$ as part of the connection.

\subsection{Symmetry algebra}  

The supercharges must transform in the fundamental representation of $SU(2)\times U(1)$ and therefore we require the following commutators
\begin{align}
&[\TT_I,\QQb_\alpha^i]=-\frac{i}{2}\QQb_\alpha^j (\sigma_I)_j^{\ i}\,, \quad [\TT_I,\QQ^\alpha_i]=\frac{i}{2}(\sigma_I)_i^{\ j}\QQ^\alpha_j\,,&\label{su2-u1-2}\\
&[\ZZ,\QQb_\alpha^i]=-i\QQb_\alpha^i\,, \quad [\ZZ,\QQ^\alpha_i]=i\QQ^\alpha_i\,,& \label{su2-u1-1}
\end{align}
where $\sigma_I$ are Pauli matrices, $\ZZ$ is the $U(1)$ generator and $\TT_I$ are the generators of $SU(2)$. The supercharges also carry a representation of $SO(3,1)$.

The generators of $SO(3,1)$, $SU(2)$, $U(1)$ and SUSY transformations --$\JJ_{ab}$, $\TT_I$, $\ZZ$, $\QQ$ and $\QQb$, respectively--, do not form a closed algebra by themselves. In fact, the anticommutator of supercharges requires enlarging the spacetime symmetry from the Lorentz group $SO(3,1)$ to the conformal group $SO(4,2)$, 
\begin{align}
 \{\QQ^\alpha_i,\QQb_\beta^j\}=&\left(\frac{1}{2}(\gamma^a)^\alpha_{\ \beta} \JJ_a-\frac{1}{2}(\Sigma^{ab})^\alpha_{\ \beta} \JJ_{ab}-\frac{1}{2}(\tilde{\gamma}^a)^\alpha_{\ \beta} \KK_a+\frac{1}{2}(\gamma^5)^\alpha_{\ \beta} \DD\right)\delta^j_i \nonumber 
 \\ &+\delta^\alpha_\beta\left(-i(\sigma^I)_i^{\ j}\TT_I-\frac{i}{4}\delta^j_i \ZZ\right)\,.\label{QQacomm}
\end{align}
where $\tilde{\gamma}_a=-\gamma_5\gamma_a$ and $\gamma_5=i\gamma^0 \gamma^1 \gamma^2 \gamma^3$. The resulting algebra includes AdS translations ($\JJ_a$), special conformal transformations ($\KK_a$) and a dilation $\DD$. These additional generators, together with the previous ones form the superalgebra $su(2,2|2)$ (see Appendix A). 

 Comparing with conformal supergravity it can be observed that the superalgebra above does not possess matching numbers of bosonic and fermionic generators and this missmatch translates to the number of bosonic and fermionic fields in our model, as we deviate from conformal SUGRA models by not demanding the so called ``conventional constrains''. For a review, see \cite{Fradkin:1985am}. This is in line with the fact that the sought gauge symmetry is just \gaugeG that does not require any additional constraint at the classical level. Additionally, there is no obstruction to make contact with conventional gauge theories because, apart from the Higgs potential and bare masses, those theories are invariant under the bigger group of transformations that includes dilations and special conformal transformations.

\subsection{Connection and curvature}\label{connandcurv} 
Using the generators of the $su(2,2|2)$ superalgebra we define the gauge connection as
\begin{equation}\label{superconnection}
\AAA= \Omega+\AAA_\psi\ \,,
\end{equation}
where $\AAA_\psi$ is given in \eqref{Afermion} and $\Omega$ contains all the bosonic fields,
\begin{align}
\Omega=& \half \omega^{ab}\JJ_{ab}+f^a \JJ_a+g^a\KK_a+h\DD +A^I \TT_I+A\ZZ\ \,.
\end{align}

The field strength, $\FF=d\AAA+\AAA \wedge \AAA$, is then given by
\begin{align}
\FF=&\half \calF^{ab} \JJ_{ab}+\calF^a \JJ_a +\calG^a \KK_a +\calH \DD\nonumber\\
&+\calF^I \TT_I +\calF \ZZ \nonumber +\QQb_\alpha^i \Psi_i^\alpha +\overline{\Psi}_\alpha^i \QQ_i^\alpha\,,
\end{align}
where the generalized curvatures are
\begin{align}
 \calF^{ab}&=\mathcal{R}^{ab}-\overline{\psi}^i\se\Sigma^{ab}\se\psi_i\,\label{Fab}\\
 \calF^a&=Df^a+ g^a h+\frac{1}{2}\overline{\psi}^i\se\gamma^a\se\psi_i\,,\label{Dfa}\\
 \calG^a&=Dg^a+ f^ah-\half \overline{\psi}^i\se\tilde{\gamma}^a\se\psi_i\,,\label{Dga}\\
 \calH  &=H+ f^ag_a+\half \overline{\psi}^i\se\gamma_5\se\psi_i\,,\\
 \calF^I&= F^I-i\overline{\psi}^i\se(\sigma^I)_i^{\ j}\se\psi_j\,,\\
 \calF &=F-\frac{i}{4}\overline{\psi}^i\se\se\psi_i\,,\label{F}\\
 \Psi_i^\alpha &=D_\Omega(\se \psi_i)^\alpha\,,\label{Fi}\\
 \overline{\Psi}_\alpha^i&=-(\overline{\psi}^i\se)_\alpha\overleftarrow{D}_\Omega\,,\label{Fibar}
\end{align}
and
\begin{align}
H&=dh\,,\\
\mathcal{R}^{ab}&=R^{ab}+ f^af^b-g^ag^b\,,\\
 R^{ab}&=d\omega^{ab}+\omega^a_{\ c}\omega^{cb}\,,\\
 F^I&=dA^I+\half \epsilon^I_{\ JK}A^J A^K\,,\\
 F&=dA\,.
\end{align}
In (\ref{Dfa}) and (\ref{Dga}) $D$ is the Lorentz covariant derivative (e.g., $DV^a=dV^a+\omega^a_{\ b}V^b$), while in \eqref{Fi}, \eqref{Fibar} $D_\Omega$ denotes the conformal covariant derivative acting on the fermion,
\begin{equation}
 D_\Omega \chi_i=\hat{D}\chi_i+\frac{1}{2}f^a \gamma_a\chi_i +\half g^a\tilde{\gamma}_a\chi_i+\half h \gamma_5\chi_i\,.
\end{equation}
where $\chi= \se \psi$ and we defined the covariant derivative $\hat{D}$ for the $SO(1,3)\times SU(2)\times U(1)$ connection, that is
\begin{align}
 (\hat{D})^{\alpha j}_{i \beta}&=\delta^j_i \delta^\alpha_\beta d +\half \omega^{ab}\delta^j_i(\Sigma_{ab})^\alpha_{\ \beta}-\frac{i}{2}A^I(\sigma_I)_i^{\ j}\delta^\alpha_\beta-iA\delta^j_i \delta^\alpha_\beta\,,\label{Dcov}\\
 (\overleftarrow{\hat{D}})^{\alpha j}_{i \beta}&=\overleftarrow{d}\delta^j_i \delta^\alpha_\beta -\half \omega^{ab}\delta^j_i(\Sigma_{ab})^\alpha_{\ \beta}+\frac{i}{2}A^I(\sigma_I)_i^{\ j}\delta^\alpha_\beta+iA\delta^j_i \delta^\alpha_\beta\,.
\end{align}
The left-acting exterior derivative of an $m$-form $\alpha^m$ is $\alpha^m\overleftarrow{d}=(-1)^m d\alpha^m$.

\subsection{Action principle} \label{actionsection} 
An action principle in 4 dimensions for the fields in the connection \eqref{superconnection} is
\begin{equation}\label{action}
 S[\omega^a{}_b,f^a,g^a,h,A^I,A, \psi]=-\int \langle \FF \circledast \FF\rangle \,,
\end{equation}
where the duality operator $\circledast$ is an involution generalizing the Hodge-$\ast$ operator and $\langle \cdots \rangle$ denotes an invariant symmetrized supertrace. Its definition is an important ingredient in our construction. This involution generalizes the analogous operator used by MacDowell and Mansouri in \cite{MacDowell:1977jt} to construct the action principle of simple SUGRA (Einstein-Hilbert plus Rarita-Schwinger terms), where the $Osp(1|4)$ SUSY breaks down to the Lorentz group. Analogously, the $\circledast$ involution breaks the $SU(2,2|2)$ gauge symmetry down to $SO(1,3)\times SU(2)\times U(1)$. In the case of \cite{MacDowell:1977jt} this explicit symmetry breaking can be traced to the fact that there exist no $SO(3,2)$-invariant tensors in four dimensions. In our case, we face a similar situation because in four dimensions there are no $SO(4,2)$-invariant tensors either.

One would like define $\circledast$ in such a way that $\langle \FF \circledast \FF \rangle$ contains the Dirac --or Weyl-- kinetic term and that any possible second order terms be confined to a boundary term,  $\partial \bar{\psi}\partial \psi \sim \partial \Omega$. In the gravity and the internal symmetry sectors, on the other hand, one would expect to reproduce the Einstein-Hilbert and the Yang-Mills terms, respectively. Naturally, when acting on the curvature components along the internal symmetries, we take $\circledast = \ast$ and, in order to restrict the possible choices for $\circledast$, one should require that $\circledast^2= -1$ so that it defines an automorphism of the $\mathfrak{su}(2,2|2)$-valued forms. 

As shown in \cite{APZ}, the operator that does the right job for the fermion and gravitational sectors, reduces to multiplication of the generators by $i\gamma_5$, where $\gamma_5$ is naturally embedded in the superalgebra. Multiplying the generators $\JJ_{a}$ and $\KK_{a}$ by $i\gamma_5$ interchanges them, and is therefore an automorphism of the algebra. However, the action of $\circledast$ along dilatations cannot be implemented in the same way because $i\gamma_5 \DD$ is not traceless and therefore it does not produce an element of the same algebra. Hence we take $\circledast H = \ast H$. In appendix (\ref{App2}) we also discuss the dynamical implications of choosing $\circledast = i \gf$. 

In \cite{APZ}, it is shown that the above choice of $\circledast$ turns $f^a$ into an auxiliary field and now the same happens for both fields $f^a$ and $g^a$.\footnote{The fact that these fields have no kinetic terms can be seen from the super traces $\langle  \JJ_{a}(i\gamma_5)\JJ_{b}\rangle=0$, $\langle  \KK_{a}(i\gamma_5)\KK_{b}\rangle=0$ and $\langle  \JJ_{a}(i\gamma_5)\KK_{b}\rangle=-\langle \KK_{a}(i\gamma_5)\JJ_{b}\rangle \propto \eta_{ab}$, which implies that the corresponding curvatures (\ref{Dfa}) and (\ref{Dga}) will not appear in the action.} This is a welcome feature of the model because $f^a$ and $g^a$ seem to play no dynamical roles in conventional theories such as General Relativity or the Standard Model. As will be seen below, the elimination of the auxiliary fields $f^a$ and $g^a$ allow us to obtain General Relativity (GR) plus Dirac or a chiral model for matter.
 
 The above considerations lead to the following definition of $\circledast$ acting on the curvature:
\begin{align}
\circledast \FF= i\Gamma_5 & \left(\half \calF^{ab} \JJ_{ab}+\calF^a \JJ_a +\calG^a \KK_a\right) +\ast\left(\calH \DD+\calF^I \TT_I +\calF \ZZ \right) \nonumber \\ & -i\QQb \gamma_5 \Psi -i\overline{\Psi}\gamma_5 \QQ \,, \label{dualoperator2-2}
\end{align}
where $(\Gamma_5)^A_{\phantom{A}B}$ is in the upper left block of the $6\times 6$ matrix representation of $su(2,2|2)$ (see Appendix 1). In (\ref{dualoperator2-2}) there is a freedom in the choice of sign for the dual operator acting on the fermion as it can be $(\pm i \gf)$ since the Dirac kinetic term is not positive definite \cite{ACDAT}. We will see below that choosing the negative sign produces a cancellation of Pauli-like couplings between the fermions and internal gauge fields.

\subsection{Effective Lagrangian}
The resulting Lagrangian is given by
\begin{align}
\calL=&\frac{1}{4}\epsilon_{abcd}\calF^{ab}\calF^{cd} -\half \calF^I\ast \calF^I -4\calF \ast\calF-\calH \ast \calH -2i \overline{\Psi} \gamma_5 \Psi\,. \label{L2}
\end{align}
 
 Let us note that $\gamma_5$ introduces a grading among the bosonic generators and therefore it is convenient to split the connection $\Omega$ as
\begin{equation}
 \Omega=\Omega^++\Omega^-\,,
\end{equation}
where ($\Omega^-$) $\Omega^+$ contains generators that (anti-)commute with $\gamma_5$. In the spinorial representation these connection components are
\begin{align}
 \Omega^+=&\frac{1}{2}\Sigma_{ab}\omega^{ab}-\frac{i}{2} \sigma_I A^I-iA+\half\gamma_5 h\,,\\
 \Omega^-=&\frac{1}{2} \gamma_a f^a+\half   \tilde{\gamma}_a g^a\,.\label{Omega-}
\end{align}

A key point is that the kinetic term for the fermion comes from $\overline{\Psi} \gf\Psi=-(\psibar \se) \lD_\Omega \gf D_\Omega \se\psi$
. In fact, the term $\overline{\Psi} \gf\Psi$ contains a term with one derivative of $\psi$, plus an algebraic bilinear in $\psi$ and a boundary term,
\begin{align}
\overline{\Psi} \gf \Psi=&\frac{i}{2}K(\overline{\chi},\chi) + \overline{\chi} \Omega^- \gf \Omega^- \chi+\overline{\chi}\gf (D^+)^2\chi + d[\overline{\chi}\gf D^+\chi]\, ,\label{3rdline}
\end{align}
where $\chi:=\se \psi$, $\overline{\chi}:=\psibar \se$,
\begin{align}
K(\overline{\chi},\chi)=&2i \overline{\chi}[\overleftarrow{D} \gamma_5 \Omega^- + \gamma_5\Omega^- D]\chi\, , \label{Kchichi1}
\end{align}
and
\begin{align}
(D^+)^2=\half \Sigma_{ab} R^{ab}-\frac{i}{2} \sigma_I F^I-iF+\half \gf H\,,\\
D=d+W\,, \quad W=\frac{1}{2}\, \Sigma_{ab} \omega^{ab}-\frac{i}{2} \sigma_I A^I -iA \,.\label{covder}
\end{align}

The term $K(\overline{\chi},\chi)$ contains what will be later identified as the Dirac kinetic term in curved space (see section \ref{EOM}), plus a piece that contains a coupling of the torsion with the vector and axial-vector current as well. That is to say,
\begin{align}\label{Kchichi2}
K(\overline{\chi},\chi)=&K_{\text{Dirac}}+K_\text{torsion}\,,
\end{align}
where
\begin{equation}\label{KDirac}
 K_{\text{Dirac}}=2i\psibar (\lD \gf\se \Omega^- \se + \gf \se\Omega^- \se D)\psi\,, 
\end{equation}
\begin{equation}\label{Ktorsion}
 K_\text{torsion}=2i\psibar\gf(\sT \Omega^- \se -\se\Omega^- \sT)\psi\,,
\end{equation}
and $\sT=T^a \gamma_a$, $T^a=De^a$.

As seen from (\ref{covder}) and (\ref{KDirac}), the field $h$ drops out from the covariant derivative $D$. The result is an \gaugeG gauge invariant kinetic term for the fermion, which is not invariant under dilations \footnote{The kinetic term \eqref{KDirac} is not even globally invariant under rigid dilations, $\psi \to e^{a\gamma_5} \psi$.}, see discussion below.

The third and fourth terms of the r.h.s. of (\ref{3rdline}) are effective mass-like and Pauli-like coupling terms between the fermionic currents and the curvatures associated to $\Omega^-$ and $\Omega^+$. The choice of $\circledast$ with the signs in (\ref{dualoperator2-2}) makes those Pauli-like couplings cancel with similar terms coming from the expansion of $-\langle \FF \circledast \FF \rangle$ along $\calF^I$, $\calF$ and $\calH$. The full quadratic action for the matter field becomes
\begin{align}\label{matquad}
\calL_\text{mat}(\psi^2)= K(\overline{\chi},\chi) +2i f\cdot g\psibar \se \se \psi + \psibar \mathfrak{R} \psi -2id[\overline{\chi}\gf D^+\chi]\,,
\end{align}
Note that the nonminimal gravitational coupling $\psibar \mathfrak{R} \psi = -4i \psibar \se \gamma_5 \slashed{R} \se \psi$ becomes and effective mass term $\sim \dV \ m_\text{eff}\psibar \psi$ for constant Lorentz curvature, e.g., in (anti-)de Sitter.

The Lagrangian \eqref{L2} also contains the standard kinetic terms for $SO(3,1)$, $SU(2)$ and two $U(1)$ gauge fields,
\begin{align}
 \calL_\text{gauge}= \frac{1}{4}\epsilon_{abcd}\mathcal{R}^{ab}\mathcal{R}^{cd}-\half F^I \ast F^I -4F\ast F-H \ast H\,.\label{Lgauge2}
\end{align}
Here $\ast$ in the Maxwell and Yang-Mills terms is the standard Hodge operator defined on a spacetime endowed with a metric $g_{\mu \nu}=\eta_{ab}e^a_\mu e^b_\nu$. There are no kinetic terms for the fields $f^a$ and $g^a$, which matches the fact that the action is not gauge invariant under transformations generated by $\JJ_a$ or $\KK_a$.

The Lagrangian also contains a Nambu--Jona-Lasinio (NJL) term,
\begin{equation}
 \calL_\text{mat} (\psi^4) =12 |e| d^4 x \left[(\psibar \psi)^2+(\psibar \gf\psi)^2\right]\,, 
\end{equation}
and the complete Lagrangian in \eqref{L2} reads
\begin{equation} \label{L3}
\calL = \calL_\text{gauge} + \calL_\text{mat}(\psi^2) + \calL_\text{mat}(\psi^4)\;.
\end{equation}

\subsection{Dilation symmetry}
As a result the $SU(2) \times U(1)$ sector is described by standard Yang-Mills and Maxwell terms minimally coupled to the spin-1/2 field. The invariance under local dilations, however, has some subtleties. The first four terms in (\ref{L2}) are indeed invariant under transformations generated by $G=\lambda(x) \DD$,
\begin{align}
\delta h =& d\lambda\,,\\                       
\delta f^a =& \lambda g^a\,,\\
\delta g^a =& \lambda f^a\,,\\
\delta \chi =& -\frac{\lambda}{2} \gf \chi\,,\\
\delta \overline{\chi} =& \frac{\lambda}{2} \overline{\chi} \gf\,,
\end{align}

The last two terms explicitly break dilation symmetry of the action. This may be counter intuitive since $\DD \sim \gf$ in the spinorial representation, but explicit calculation shows
\begin{align}
 \delta \Psi=&  -\frac{\lambda}{2} \gf \Psi+2\lambda \gf \Omega^- \chi\,,\\
 \delta \overline{\Psi}=&  \frac{\lambda}{2} \overline{\Psi}\gf+2\lambda \overline{\chi}\gf \Omega^- \,,
\end{align}
and therefore the Lagrangian changes by
\begin{equation}
 \delta \calL = 4 i \lambda \overline{\chi} (\lD \Omega^-+\Omega^- D)\chi\;, 
\end{equation}
which, in general, is not a total derivative. This is in fact proportional to a linear combination of the field equations for the spinor and is therefore an on-shell invariance that is likely to be anomalous. On the other hand,  since $h$ drops out from the covariant derivative $D$ in the Lagrangian --it only occurs in a boundary term in \eqref{L3}--, the fermionic Lagrangian is not expected to reflect the symmetry generated by $\DD$. From the fact that the kinetic term for $h$ is just a Maxwell term, one can conclude that $h$ becomes a sterile abelian boson field that can be interpreted as a hidden photon that contributes to the energy content of a hidden sector. If a kinetic mixing term between the $U(1)$ gauge field and $h$ were present in the action, the propagating states would mix producing interesting phenomenology as discussed in \cite{Holdom:1985ag}.

\section{Physical contents} \label{physicalcontent} 
Let us now examine the physical content of the theory as defined by \eqref{L3}. Since the gauge sector is given by standard $SU(2) \times U(1)$ Yang-Mills-Maxwell terms (plus the corresponding minimal couplings), we will focus on the derivation of the gravitational and matter sectors.

The absence of kinetic terms for $f^a$ and $g^a$ in \eqref{L2} and the fact that they appear only algebraically in the Lagrangian indicates that these are auxiliary fields. Therefore, they can be substituted from their own field equations back into the action, without losing dynamical information. The fields $f^a$ and $g^a$ appear in the gravitational Lagrangian,
\begin{equation} \label{LG}
\frac{1}{4}\epsilon_{abcd}\mathcal{R}^{ab}\mathcal{R}^{cd} =\frac{1}{4}\epsilon_{abcd} (R^{ab}+ f^af^b-g^ag^b)(R^{cd}+ f^cf^d-g^cg^d)\, , 
\end{equation}
and in the kinetic term of the matter Lagrangian
\begin{equation} \label{LD}
\left. \calL_\text{mat} \right|_{f^a,g^a}= 2i (\psibar \se)[\overleftarrow{D} \gamma_5 \Omega^- + \gamma_5 \Omega^- D](\se \psi) +2i f\cdot g\psibar \se \se \psi\,, 
\end{equation}
where $\Omega^-= \frac{1}{2} \gamma_a f^a+\half   \tilde{\gamma}_a g^a$. From these expressions it is apparent that although the field equations for $f^a$ and $g^a$ are algebraic and can in principle be solved for them, those solutions would depend on the configurations of $\omega^{ab}$ and $\psi$, which in turn depend on the $SU(2)$ and $U(1)$ gauge fields. Therefore, solving for the auxiliary fields $f^a$ and $g^a$ can be best understood on a region around a certain vacuum. A sensible vacuum consists of a state devoid of matter and gauge fields with the largest possible symmetry, which in the present case would be \gaugeG.

\subsection{(A)dS vacuum sector} 
Consider the purely bosonic sector ($\psi=0$) where the field equations for $f^a$ and $g^a$ are
\begin{align}
  \left.  \frac{\delta \calL}{\delta f^a}\right|_{\psi=0}
=&\epsilon_{abcd}f^b(R^{cd}+f^c f^d-g^c g^d)\,,\label{eqfbosonic}\\
  \left. \frac{\delta \calL}{\delta g^a}\right|_{\psi=0}
=&-\epsilon_{abcd}g^b(R^{cd}+f^c f^d-g^c g^d)\,. \label{eqgbosonic}
\end{align}

These equations admit a maximally symmetric, constant Lorentz curvature solution, like the AdS or dS vacua,
\begin{equation} \label{(A)dS}
R^{ab} \pm \ell^{-2} e^a e^b =0,
\end{equation}
where the cosmological constant is $\Lambda \varpropto \mp \ell^{-2}$. The AdS vacuum is also an interesting solution because it is invariant under $SO(3,1)\times SU(2) \times U(1)$ and it admits a maximal number of Killing spinors, i.e., it is a BPS state with the maximal number of supersymmetries \cite{ATZ}.

Compatibility between (\ref{eqfbosonic}), (\ref{eqgbosonic}) and \eqref{(A)dS} requires 
\begin{equation}
f^af^b - g^ag^b = \pm \ell^{-2} e^a e^b , 
\end{equation}
and therefore this solution can be parametrized, in the AdS (+) case, as
\begin{align} \label{fgfixing}
f^a = \frac{ e^a}{\ell} \cosh{\lambda} \; , \qquad
g^a = \frac{ e^a}{\ell} \sinh{\lambda} \;\, ,
\end{align}
where $\lambda$ is some real parameter.\footnote{For dS, the roles of $f^a$ and $g^a$ must be exchanged.\label{foot}} From now on we will refer to the sector\eqref{(A)dS} with \eqref{fgfixing} as the GR sector. Substituting back in the action the solutions for the auxiliary fields produces an equivalent dynamical system --a consequence of the implicit function theorem for functional equations \cite{Henneaux-Teitelboim}.

The effective action is now a functional of the standard fields of first order gravity ($e^a$, $\omega^{ab}$), the internal $SU(2)$ and $U(1)$ gauge connections ($A^I$, $A$), and the spin-1/2 matter multiplet ($\psi$).

\subsubsection{Gravitional effective action} \label{gravitysection}%
In the vacuum sector (\ref{fgfixing}), the effective gravitational Lagrangian contains the Einstein-Hilbert and cosmological constant terms plus the Euler density (surface term),
\begin{align} \label{GR}
\frac{1}{4}\epsilon_{abcd}\mathcal{R}^{ab}\mathcal{R}^{cd} =\frac{1}{4}\epsilon_{abcd} (R^{ab} \pm \ell^{-2}e^a e^b)(R^{cd} \pm \ell^{-2}e^c e^d)\, , 
\end{align}
where (+) corresponds to AdS and ($-$) to dS. Hence, dropping the Euler term, the effective gravitational action is given by
\begin{align}
 \frac{1}{4}\int\epsilon_{abcd}\mathcal{R}^{ab}\mathcal{R}^{cd} &= \pm \frac{1}{2\ell^2} \int \epsilon_{abcd} (R^{ab}e^c e^d \pm \frac{1}{2\ell^2} e^a e^b e^c e^d) \nonumber\\
 &=\pm\frac{1}{\ell^2} \int d^4x |e| \left(R \pm \frac{3}{\ell^2}\right). \label{EH1}
\end{align}
This matches the standard form of the Einstein-Hilbert action,
\begin{align}
 I_{EH} = \frac{1}{16\pi G_N} \int d^4x |e| (R-2\Lambda)\,,\label{gravityaction}
\end{align}
provided one identifies\footnote{Here we used $\epsilon_{abcd}R^{ab}e^c e^d=2\dV R\,$, $\epsilon_{abcd}e^a e^b e^c e^d=4! \dV$ and we follow the sign conventions of \cite{MTW}.}
\begin{equation} \label{G-CC}
\pm \frac{1}{\ell^2}= \frac{1}{16\pi G_N}=\frac{M_P^2}{2}\,, \qquad \Lambda = \mp\frac{3}{\ell^2}= \mp \frac{3}{2}M_P^2.  
\end{equation}

Note that in \eqref{G-CC} both signs are allowed (see footnote \ref{foot}), provided that in the dS case the would-be wrong sign in front of the action \eqref{EH1} is compensated by the replacement $\Gamma_5 \rightarrow -\Gamma_5$ in (\ref{dualoperator2-2}). The fact that both signs of the cosmological constant are admissible in the model can be traced to the possibility of choosing the underlying spacetime symmetry as AdS (\ref{ads2}) or dS (\ref{ds}) as subalgebras of the conformal algebra. If instead of the superconformal algebra we had started from superAdS, in order to change the sign of the cosmological constant we would have been forced to introduce an $i$-factor in $\Omega^-$, which would be inconsistent with unitarity of the fermion sector and would not allow for chiral models without the introduction of mirror fermions \cite{Chamseddine:2013hwa}.

Varying the Lagrangian \eqref{L3}, for the choice \eqref{fgfixing}, with respect to the vierbein reproduces the Einstein equations with the right stress-energy tensor source generated by the gauge and matter fields.

\subsubsection{Matter Lagrangian} \label{mattersection}
In the sector (\ref{fgfixing}), the fermionic Lagrangian in \eqref{L3} takes the form 
\begin{equation}\label{Dirac'}
 \calL_{\mbox{mat}}= 2i \psibar (\lD \gf  \se \Omega^- \se + \gf \se \Omega^- \se D)\psi + K_\text{torsion} + \calL_\text{mat}(\psi^4) \, ,
\end{equation}
with
\begin{align} \nonumber
\Omega^- &=(2 \ell)^{-1}(\cosh \lambda + \gamma_5 \sinh \lambda)\se \; =\; (2 \ell)^{-1}e^{\lambda \gamma_5}\se\;, \\
& =\se (\alpha \Pi_++\beta \Pi_-)\,,\label{Omega-GRsector}
\end{align}
where $\Pi_{\pm}=(1\pm \gamma_5)/2$ are the chiral projectors, and
\begin{equation}
\alpha=\frac{\cosh \lambda - \sinh \lambda}{2 \ell}\,, \quad \beta=\frac{\cosh \lambda + \sinh \lambda}{2 \ell}\,.
\end{equation}

The kinetic term in \eqref{Dirac'} is the Dirac Lagrangian for a linear combination of right- and left-handed fermions ($\psi_+$, $\psi_-$). The Lagrangian for chiral or anti-chiral fermions is obtained for $\Omega^-=1/2(1 \pm \gamma_5)\se$, which corresponds to $f^a=\pm g^a$. These cases can only be reached in the vanishing cosmological constant limit $\ell\to \infty$ and $\lambda \to \pm \infty$ with $e^{|\lambda|}/\ell^2=$ constant.\\

\subsection*{Non chiral fermions: $\alpha \neq 0\neq \beta$ ($|\lambda|< \infty $)}
For finite $\lambda$, the first term in \eqref{Dirac'} describes a Dirac field minimally coupled to the \gaugeG gauge connection,
\begin{align}
K_\text{Dirac} =&-\half |e| d^4 x \left[ \psibar'(\slashed{D}-\overleftarrow{\slashed{D}})\psi'\right]\,, \qquad \psi'=(\psi'_-,\psi'_+)\, ,
\end{align}
where we have defined the physical Weyl spinors by
\begin{equation}\label{physicalspinor1}
 \psi_-'=\sqrt{24\alpha} \psi_-\,, \quad \psi_+'=\sqrt{24\beta} \psi_+\,.
\end{equation}

The covariant slashed derivatives are, in agreement with (\ref{covder}), given by 
\begin{equation}
 \slashed{D}=\gamma^\mu (\partial_\mu+W_\mu)\,, \quad \overleftarrow{\slashed{D}}= (\overleftarrow{\partial}_\mu-W_\mu)\gamma^\mu\,.
\end{equation}

As a consequence of \eqref{fgfixing} and the the fact that the spinor enters in the action through the combination $\se \psi$, torsion couples to the axial current of the physical spinor, in addition to the usual minimal couplings of fermions in curved space,
\begin{align}
K_\text{torsion}=\frac{2i}{3} T \cdot e \ e^a(\psibar'\gamma_5 \gamma_a\psi')\,.
\end{align}

In the sector defined by \eqref{fgfixing} the term $2i f.g \psibar \se \se \psi$ vanishes and the non minimal coupling to the curvature in \eqref{matquad} reduces, in the AdS vacuum, to an ``effective mass" for the fermion
\begin{equation}\label{meff}
 \psibar \mathfrak{R} \psi= -\frac{4i}{4!\ell^2\sqrt{\alpha \beta}}\psibar'\gamma_5 \slashed{R}\se\se \psi' = \frac{8}{\ell}|e| d^4x \psibar' \psi' = m_\text{eff} \psibar' \psi' \,.
\end{equation}

So far, the fermionic Lagragian if expressed in terms of the Dirac fermion $\psi'$ has no trace of $\lambda$, which means that this is an irrelevant parameter that can be safely set to zero.\footnote{As can be easily seen, this is also true of the NJL terms.}

Comparing \eqref{meff} with \eqref{G-CC}, the scale of the effective mass is of the order of the Planck mass. This is a consequence of the tight relation between all the parameters which is typical of supersymmetric theories and in particular of conformal SUGRA \cite{Fradkin:1985am}. However, by changing the Lagrangian by an overall constant, 
\begin{equation}\label{Lscaling}
\calL \rightarrow \xi \calL\,, 
\end{equation}
modifies Newton's constant and therefore the value of the cosmological constant in terms of the Planck mass gets redefined
\begin{equation}\label{lambdascaled}
 \Lambda = -\frac{3}{2\xi} M_P^2\,.
\end{equation} \\
For instance, in a $\Lambda$CDM cosmological scenario one would expect $\xi \sim 10^{120}$.\\

\subsection*{Chiral fermions: $\alpha \ne 0$ and $\beta = 0$ ($|\lambda|=\infty$)}
Let us now consider the ``chiral sector'' of the theory. From (\ref{KDirac}) and (\ref{Omega-GRsector}) we see that $\Omega^-$ ($\sim \se \Pi_+$) projects out the right handed fermion from the kinetic term of the matter action (for $\alpha= 0$ and $\beta \ne 0$ the left handed fermion gets projected out), resulting in a Weyl kinetic term for the leftt handed chiral spinor only,
\begin{align}
K_\text{Weyl} =&-\half |e| d^4 x \left[ \psibar'_-(\slashed{D}-\overleftarrow{\slashed{D}})\psi'_-\right]\,.
\end{align}
Therefore the right handed chiral spinor does not propagate, becoming an auxiliary field in this limit. As expected, the mass term, given by (\ref{meff}), vanishes in the chiral sector. The NJL term is the only part of the action containing $\psi_+$, see (\ref{NJL}), and on shell $\psi_+$ is forced to vanish. 

As pointed out above, the chiral limit is obtained for $\lambda \to 0$ and simultaneously $\ell \to \infty$, which is rather singular because not only the cosmological constant vanishes, but the entire gravitational action goes away. This may be interpreted in the sense that the resulting theory describes a chiral left handed fermion coupled to a $SU(2)\times U(1)$ plus an extra abelian sterile boson in a non dynamical spacetime background, a completely renormalizable model.

\subsubsection*{NJL coupling}

The quartic fermionic terms from the first four contributions on the r.h.s. of (\ref{L2}) give the NJL coupling. The contributions coming from the Abelian subgroups are
\begin{align}\label{T4-abel}
 T^{\calF \ast \calF }(\psi^4) + T^{\calH \ast \calH }(\psi^4) =
 \frac{1}{4}(\psibar \se \se \psi)(\psibar \ast(\se \se) \psi) +\frac{1}{4}(\psibar \se\gf\se \psi)(\psibar \ast(\se \gf\se) \psi). 
\end{align}
In four spacetime dimensions, the Hodge-$\ast$ operation and multiplication by $i\gamma_5$ have the same effect on $\se \se$: $\ast \se\se = i \gamma_5\se\se$. Using this, and the Fierz identity $\phi^{ab5}\phi_{ab5}= \phi^{ab}\phi_{ab}$, it is easy to show that \eqref{T4-abel} vanishes identically.

The Lorentz and $SU(2)$ contributions are
\begin{align}\label{LorentzNJL}
 T^{\calF^{ab} \calF^{cd} }(\psi^4)& =\frac{1}{4}\epsilon_{abcd}(\psibar \se \Sigma^{ab} \se \psi)(\psibar \se \Sigma^{cd} \se \psi), \\
\label{SU2NJL}
 T^{\calF^I \ast \calF^I }(\psi^4)& =\frac{1}{2}(\psibar \se \sigma^I \se \psi)(\psibar \ast(\se \se)\sigma^I \psi).
\end{align}
By direct computation of (\ref{LorentzNJL}) and Using the Fierz identity $\phi^{abI}\phi_{abI}=6(\phi^2+\phi_5^2)$ in (\ref{SU2NJL}) we arrive at\footnote{(\ref{NJL}) vanishes for the de Sitter case.}
\begin{equation}\label{NJL}
 T(\psi^4)=12|e|d^4 x (\phi^2+\phi_5^2)\,.
\end{equation}
In terms of the rescaled physical spinor we get
\begin{align}
 T(\psi^4)=\frac{\ell^2}{12}|e|d^4 x (\phi' {}^2+\phi'_5 {}^2)\,,
\end{align}
where we can see that the quartic fermion term is suppressed by a $M_P^{-2}$ coupling,
\begin{equation}\label{gNJL}
 g_{NJL}=\frac{1}{6M_P^2}\,.
\end{equation}\\

\subsection{Effective bare coupling constants} 
The model also predicts values for the bare gauge coupling constants $g_{U(1)}$ and $g_{SU(2)}$. Since the $SU(2)$ field has the canonical normalization in (\ref{L2}) we can identify $g_{SU(2)}=1$ directly from the covariant derivative (\ref{Dcov}). However, $g_{U(1)}$ has to be read from the covariant derivative (\ref{Dcov}) only after canonically normalizing the $U(1)$ field such that the term in (\ref{L2}) plus the kinetic term of the fermionic action read,
\begin{equation}
 \calL_\text{can}=-\frac{1}{2}F'\ast F'+\dV \psibar' \sD \psi'\,.
\end{equation}
in terms of the physical rescaled gauge field and physical spinor. As a result we can identify $ g_{U(1)} = \frac{1}{2\sqrt{2}}\sim 0.36 $. Both values are higher than the SM values ($g_{U(1)}^\text{SM}\approx 0.34 $ and $g_{SU(2)}^\text{SM}\approx 0.66 $), however the hierarchy $g_{SU(2)}>g_{U(1)}$ is respected. Leaving aside the cosmological constant problem, we can introduce an overall constant in the action to get the right values for two of the three coupling constants $(G_N,g_{SU(2)},g_{U(1)})$. The predicted Weinberg angle, however, remains unchanged, $\sin^2 \theta_W =g_{U(1)}^2/(g_{U(1)}^2+g_{SU(2)}^2)= 1/9$, and falls shorter than the SM value.

The introduction of an overall constant in the action to get the $\Lambda$CDM value for the cosmological constant, see (\ref{Lscaling}) and (\ref{lambdascaled}), implies the relations $\xi g_{SU(2)}^2=1=8\xi g_{U(1)}^2$ that correspond to a huge suppression of the gauge coupling constants. The mass squared parameter also gets highly suppressed by the $\xi$-parameter,
\begin{equation}
 m_\text{eff}^2=\frac{2}{3} |\Lambda |=\frac{M_P^2}{\xi}\,,
\end{equation}
while the NJL coupling (\ref{NJL}) remains unchanged.

The introduction of an overall factor in the action allows, in the chiral case, fitting one of the gauge couplings to the SM value. For instance we can take $\xi$ such that $\xi g_{SU(2)}^{SM}=1$, and therefore $g_{SU(2)}=g_{SU(2)}^{SM}$ and $g_{U(1)}=g_{SU(2)}/(2\sqrt{2})\sim 0.36 g_{SU(2)} \sim 0.23$. The prediction of a hierarchy of gauge couplings $g_{SU(2)}/g_{U(1)}\sim 2.8$ is also valid in the chiral case.

\subsection{Comments on the field equations}\label{EOM}
In the AdS sector, defined by the choice \eqref{fgfixing} and in the generic non-chiral case, the effective action reads
\begin{align} \label{Leff}
I = & \int |e| d^4x \left[ \frac{1}{16\pi G_N}(R-2\Lambda) -\frac{1}{4} F^{I\,\mu \nu}\; F^I_{\mu \nu}- \frac{1}{4}F^{\mu \nu}F_{\mu \nu} -\frac{1}{4}H^{\mu \nu}H_{\mu \nu} \right. \nonumber \\
& \left. -\half  \psibar'(\slashed{D} + \overleftarrow{\slashed{D}})\psi' + \frac{2}{3} \tilde{T}^\mu (\psibar i\gamma_5 \gamma_\mu \psi) + \frac{\ell^2}{12}\left[(\psibar \psi)^2 + (\psibar \gamma_5 \psi)^2 \right] \right] \, ,
\end{align}
where $D$ is the covariant derivative for the \gaugeG gauge connection, $\tilde{T}^\mu=\epsilon^{\mu \nu \lambda \rho} T_{\nu \lambda \rho}$, and we have dropped the prime from the fermionic field. Apart from the torsional coupling and the NJL term, this action is a standard Einstein-Dirac-Maxwell-$SU(2)$ system.

Varying with respect to the metric (or the vielbein) yields Einstein's equations with a stress-energy tensor produced by the gauge fields and the Dirac field (see Appendix C). Varying with respect to the Lorentz connection gives an equation that expresses the torsion as a bilinear of the fermionic field. Since this equation is algebraic in the connection, it can be substituted back in the action, giving an additional contribution to the NJL term.

The field equation for the Dirac field contains, together with the Dirac operator minimally coupled to the internal gauge connection and to the spacetime geometry, a cubic tern coming from the NJL piece. This term is the one that produces chiral symmetry breaking and the generation of a mass gap in the NJL model of superconductivity \cite{NJL, Klevansky, Vogl:1991qt}.

Finally, the internal gauge connections $A$ and $A^I$ satisfy the standard Maxwell and Yang-Mills equations in curved spacetime sourced by the standard matter currents.

\section{Summary and Outlook}\label{summary}

We have presented a bottom-up construction of a model with broken SUSY obtained by gauging a conformal superalgebra that includes local Lorentz, $U(1)$ and $SU(2)$ internal symmetries. The coupling of matter fields is achieved by including matter fermions and gauge bosons in the same super  connection for the superalgebra, using the ``matter ansatz'' (see eq. (1.3)). As a consequence, fermions are in the fundamental representation of the gauge group, there is no matching between bosonic and fermionic degrees of freedom and no pairing of particle states.

The vacuum of the theory, defined by vanishing gauge curvature configurations, $\FF=0$, is invariant under the full superconformal symmetry with supersymmetry parameter satisfying $\slashed{\nabla}\epsilon=0$. The field contents of the effective theory constructed around this vacuum consists of an $SU(2)\otimes U(1)$ Yang-Mills theory plus gravity and matter, described by a Dirac field minimally coupled to the gauge fields with NJL self-interactions.

The propagating degrees of freedom of the resulting theory are the same as in a non supersymmetric system of spin-1/2 matter minimally coupled to gravity and charged with respect to the internal gauge fields described by a Yang-Mills theory. The analysis leading to this conclusion is similar to that in the 2+1 dimensional Chern-Simons uSUSY model, where the breaking of local SUSY and scale invariance --as in the present case-- gives rise to propagating degrees of freedom in an otherwise topological theory \cite{GPZ}.

The type of spinors in the effective theory depends on the sector of the theory that emerges from the fixing of the auxiliary fields $f^a$ and $g^a$. In one sector the spinors are non chiral. In another sector, the matter fields become chiral and gravity decouples \cite{Fradkin:1985am}. Thus, the model incorporates a novel mechanism to generate chiral matter that is interesting to explore further from the point of view of unified models of gravity and leptons together with their electroweak interactions. Possible mechanisms to incorporate a Higgs doublet and other SUSY representations that could give realistic values for the gauge couplings deserve further exploration.

Some features of this model are similar to those found in conformal SUGRA, such as the prediction of the tree level coupling constants. This can serve to illustrate how to implement a model with the gauge structure of an electroweak sector, although in the present setup the predicted values for the tree level coupling constants do not come up right, and the bare cosmological constant is of the order of the Planck mass, a defect that the model shares with conformal SUGRAs \cite{Fradkin:1985am}.

As pointed out in \cite{Hooft:2014daa}, it may be interesting to look for mechanisms of spontaneous conformal symmetry breaking, although the introduction of a Higgs potential in this framework, \textit{i. e.}, as part of the gauge connection, seems somewhat unnatural. Attempts that exploit the spontaneous breaking of local scale invariance can be found for instance in \cite{Smolin:1979uz} and more recently in \cite{Edery:2006hg,deCesare:2016mml,Ghilencea:2018dqd}. Spontaneous breaking of local scaling symmetry has been also explored in the context of inflation, where it was pointed out that it can reproduce Starobinski inflation without fine tunning of parameters \cite{Barnaveli:2018dxo}. Embedding the present model in conformal supergravity \cite{Ferrara:2018wqd} might eventually provide a viable mechanism for spontaneous breaking of local scale invariance.

\section*{Acknowledgments}
Fruitful discussions with L. Andrianopoli, P. Arias, B. L. Cerchiai, R. D'Auria, L. Delage, M. P. García del Moral, J. Gomis, P. Pais, A. Restuccia and M. Trigiante, are gratefully acknowledged. This work was also partially supported by Fondecyt grant 1180368, MINEDUC-UA project ANT 1755 and by Semillero de Investigación project SEM18-02 from Universidad de Antofagasta, Chile. The Centro de Estudios Cient\'{\i}ficos (CECs) is funded by the Chilean Government through the Centers of Excellence Base Financing Program of Conicyt.

\appendix

\section{Representation of the $su(2,2|2)$ superalgebra}\label{App1}
We use the following representation of $su(2,2|2)$
\begin{align}
(\JJ_{ab})^A_{\ B}=&(\Sigma_{ab})^A_{\ B}\,, \qquad \qquad \qquad \mbox{Lorentz}\\
(\JJ_a)^A_{\ B}=&\frac{1}{2}(\gamma_a)^A_{\ B}\,, \qquad \qquad \qquad \mbox{(A)dS boosts}\\
(\KK_a)^A_{\ B}=&\frac{1}{2}(\tilde{\gamma}_a)^A_{\ B}\,, \qquad \qquad \qquad \mbox{Special conformal transf.}\\
(\DD)^A_{\ B}=&\frac{1}{2}(\gamma_5)^A_{\ B}\,, \qquad \qquad  \qquad \mbox{Dilations}\\
(\mathds{T}_I)^A_{\ B}=&-\frac{i}{2} u^{Ai}(\sigma_I)_i^{\ j} u_{jB}\,, \qquad SU(2) \\
(\mathds{Z})^A_{\ B}=&i\delta^A_\beta\delta^\beta_B+2i\delta^A_j\delta^j_B\,, \qquad \quad U(1) \,\\
(\QQ^\alpha_i)^A_{\ B}=&\delta^A_i \delta^\alpha_B\,, \qquad \qquad \qquad  \quad \; \mbox{SUSY} \\
(\QQb_\alpha^i)^A_{\ B}=&\delta^A_\alpha \delta^i_B\,, \qquad \qquad \qquad  \quad \;\; \mbox{SUSY}
\end{align}

The $\gamma$-matrices are in a $4\times 4$ Weyl representation ($\alpha, \beta,\cdots$ run from 1 to 4). The indexes of the tangent space $a,b=0,1,2,3$. Indexes in the adjoint representation of $SU(2)$ take values $I,J=1,2,3$, and in the fundamental take the values $i,j=1,2$. The $\sigma$-matrices are the usual Pauli matrices and $u^{ij}=i \sigma_2$, $u_{ij}u^{jk}=\delta_i^k$. Indexes of the representation are $A,B=1,\cdots,6$, so we have a $6\times6$ representation. We find convenient to split $A=(\alpha,i)$ and therefore it is understood that any $\gamma_a$-matrix valued in the ``$i$'' range gives a vanishing result and any $\sigma_I$-matrix valued in the ``$\alpha$'' is vanishing as well.

The Lorentz generators and AdS boosts form the AdS algebra,
\begin{align}
[\JJ_{ab},\JJ_{cd}]&= -\eta_{ac}\JJ_{bd} +\eta_{ad}\JJ_{bc}+ \eta_{bc}\JJ_{ad}-\eta_{bd}\JJ_{ac} \,\label{ads1}\\
[\JJ_a,\JJ_b]&= + \JJ_{ab}\,, \qquad
[\JJ_a,\JJ_{bc}]=\eta_{ab}\JJ_c-\eta_{ac}\JJ_b\,,\label{ads2}
\end{align}

The remaining generators of spacetime transformations satisfy
\begin{equation}
[\KK_a,\KK_b]=-\JJ_{ab}\,, \quad
[\KK_a,\JJ_{bc}]=\eta_{ab}\KK_c-\eta_{ac}\KK_b\,,\label{ds}
\end{equation}
\begin{equation} \quad 
[\DD,\JJ_a]=-\KK_a\,, \qquad
[\DD,\KK_a]=-\JJ_a\,, 
\end{equation}
which, together with the AdS generators complete the conformal algebra. The commutators between SUSY generators and those of the conformal group are
\begin{eqnarray}
&[\JJ_a,\QQb_\alpha^i]=\frac{1}{2}\QQb_\beta^i(\gamma_a)^\beta_{\ \alpha}\,, \quad [\JJ_a,\QQ^\alpha_i]=-\frac{1}{2}(\gamma_a)^\alpha_{\ \beta}\QQ^\beta_i\,,&\\
&[\JJ_{ab},\QQb_\alpha^i]=\QQb_\beta^i(\Sigma_{ab})^\beta_{\ \alpha}\,, \quad [\JJ_{ab},\QQ^\alpha_i]=-(\Sigma_{ab})^\alpha_{\ \beta}\QQ^\beta_i\,,&\\
&[\KK_a,\QQb_\alpha^i]=\frac{1}{2}\QQb_\beta^i(\tilde{\gamma}_a)^\beta_{\ \alpha}\,, \quad [\KK_a,\QQ^\alpha_i]=-\frac{1}{2}(\tilde{\gamma}_a)^\alpha_{\ \beta}\QQ^\beta_i\,,&\\
&[\DD,\QQb_\alpha^i]=\frac{1}{2}\QQb_\beta^i(\gamma_5)^\beta_{\ \alpha}\,, \quad [\DD,\QQ^\alpha_i]=-\frac{1}{2}(\gamma_5)^\alpha_{\ \beta}\QQ^\beta_i\,.&
\end{eqnarray}
Hence, the supercharges carry a representation of the full conformal algebra.
The $SU(2)$ generators commute to
\begin{equation}\label{su2-u1-4}
 [\TT_I,\TT_J]=\epsilon_{IJ}^{\ \ \ K}\TT_K\,.
\end{equation}
Finally, under the Lorentz group, $\QQ$ and $\QQb$ transform in the standard form,
\begin{align} \label{Lor}
[\JJ_{ab},\QQ^\alpha_i]=\frac{1}{2}(\Sigma_{ab})^\alpha{}_\beta \QQ^\beta_i\,, \quad [\JJ_{ab},\QQb_\alpha^i]=-\frac{1}{2}\QQb_\beta^i \left(\Sigma_{ab}\right)^\beta{}_\alpha\,,
\end{align}
where $\Sigma_{ab}=\frac{1}{4}[\gamma_a,\gamma_b]$.

The quadratic combinations that give nontrivial traces are
\begin{eqnarray}
&\langle \JJ_a \JJ_b \rangle=\eta_{ab}\,, \qquad \langle \JJ_{ab} \JJ_{cd} \rangle=-(\eta_{ac}\eta_{bd}-\eta_{bc}\eta_{ad})\,,&\\
& \langle \KK_a \KK_b \rangle=-\eta_{ab}\,, \qquad \langle \DD^2\rangle = +1\,,&\\
&\langle \TT_I \TT_J \rangle=\frac{1}{2}\delta_{IJ}\,, \qquad \langle \ZZ^2\rangle=4\,,&\\ 
&\langle \QQ^\alpha_i \QQb^j_\beta\rangle=-\delta^\alpha_\beta \delta^j_i=-\langle \QQb^j_\beta \QQ^\alpha_i\rangle\,.&
\end{eqnarray}

The operator $i \Gamma_5$ that is fundamental in getting the right kinetic terms has nontrivial traces given by
\begin{align}
 \langle i \Gamma_5 \DD \rangle &=2i\,, \\
 \langle i \Gamma_5 \JJ_{ab}\JJ_{cd}\rangle&=-\epsilon_{abcd}=\langle \JJ_{ab}i \Gamma_5\JJ_{cd}\rangle\,,\\
 \langle  \JJ_{a}i \Gamma_5\KK_{b}\rangle&=-i \eta_{ab}=-\langle \KK_{a}i \Gamma_5\JJ_{b}\rangle\,,\\
 \langle \ZZ i \Gamma_5 \DD \rangle&=-2=\langle \DD i \Gamma_5 \ZZ \rangle\,.
\end{align}

\section{Alternative choices for the dual operator}\label{App2}
A possible alternative definition of $\circledast$ could be obtained multiplying all the generators of the conformal subgroup by $i\gamma_5$. As we mentioned, this choice is not an automorphism of the algebra (in particular, the supertrace of $\gamma_5 \DD$ does not vanish). If one were to insist on this choice, then
given by
\begin{align}
\circledast \FF=&S\left(\half \calF^{ab} \JJ_{ab}+\calF^a \JJ_a +\calG^a \KK_a +\calH \DD \right)\nonumber\\
&+\ast\left(\calF^I \TT_I +\calF \ZZ \right) -i\QQb^i \gamma_5 \calF_i -i\hat{\calF}^i\gamma_5 \QQ_i\,. \label{dualoperator1}
\end{align}

The operator $S$ squares to minus the identity in the bosonic sector ocupied by the generators $\{\JJ_{ab},\JJ_a,\KK_a, \DD\}$. By using (\ref{dualoperator1}), we obtain the Lagrangian
\begin{align}
\calL=&\frac{\varepsilon_s}{4}\epsilon_{abcd}\calF^{ab}\calF^{cd} -\half \calF^I\ast \calF^I -4\calF\ast\calF \nonumber\\
& +2\varepsilon_s \calH \calF -2i \hat{\calF}^i \gamma_5 \calF_i\,. \label{L1}
\end{align}
The Lagrangian can be decomposed as $\calL =\calL_\text{gauge}+\calL_\text{matter}$ where 
\begin{align}
 \calL_\text{gauge}=&\frac{\varepsilon_s}{4}\epsilon_{abcd}\mathcal{R}^{ab}\mathcal{R}^{cd}-\half F^I \ast F^I -4F\ast F+2\varepsilon_s H F\,,\label{Lgauge1}\\
 \calL_\text{matter}=&\calL(\psi^2)+\calL(\psi^4)\,.
\end{align}
The first term in the r.h.s of (\ref{Lgauge1}) contains the Einstein-Hilbert term, the cosmological constant term and the Gauss-Bonnet term. The second term in the r.h.s of (\ref{Lgauge1}) is the Yang-Mills action for $SU(2)$ and the third term is the Maxwell action for the $U(1)$ field $A$.

In (\ref{L1}), the $h$ field becomes a Lagrange multiplier that appears in the term $\calH \calF$ only, and which imposes  restriction on $\psi$ of the form,
\begin{equation}
 d(\psibar \se \se \psi)=0\,.
\end{equation}
This equation is automatically satisfied by chiral spinors that have vanishing tensor bilinears.

A signature feature of theories constructed using (\ref{action}) is that Pauli-like couplings that may appear a priori in the action are in fact canceled as a result of the underlying SUSY principle and an appropriate choice of $\circledast$. Let us note that Pauli-like couplings are fully consistent with gauge invariance, implying a modification to the current at the classical level. However, such terms produce modifications to the dipole moments of leptons and are strongly constrained, see \cite{AbdusSalam:2019kei} and references therein.

Yet another possible choice for the dual operator is given by
\begin{align}
\circledast \FF=&S\left(\half \calF^{ab} \JJ_{ab}+\calH \DD\right)+\calF^a \JJ_a +\calG^a \KK_a\nonumber\\
& +\ast\left(\calF^I \TT_I +\calF \ZZ \right)\nonumber\\
&-i\QQb^i \gamma_5 \calF_i -i\hat{\calF}^i\gamma_5 \QQ_i\,.\label{dualoperator2}
\end{align}
This choice implies that the Lagrangian is given by (\ref{L2}) plus terms that are a priori topological,
\begin{equation}\label{topfa}
 \calF^a\calF_a\,, \quad \calG^a\calG_a\,.
\end{equation}
The topological origin of (\ref{topfa}) means that the appearance of such terms in the action will not assign independent dynamics to the fields $f^a$ and $g^a$ and therefore they are not forbidden. These terms, however, have to be analysed along the term $-2i \hat{\calF}^i \gamma_5 \calF_i$ that has the \gaugeG gauge invariance only. This means that the equations of motion coming from the variation of $f^a$ and $g^a$ in (\ref{topfa}) will impose certain restrictions on some currents of the spinor bilinears.

So far we have introduced three independent choices, (\ref{dualoperator2-2}), (\ref{dualoperator1}) and (\ref{dualoperator2}), for the dual operator that provide gauge theories with \gaugeG gauge invariance and are chiral invariant.

\section{Field Equations}
The fields of the effective theory \eqref{Leff} are $e^a_{\mu}$, $\omega^a{}_{b \mu}$, $\psi$, $\bar{\psi}$, $A^I_\mu$ and $A_\mu$. The Einstein equations obtained varying with respect to  $e^a_\mu$ read,
\begin{align}
\frac{1}{\kappa}(G^\mu{}_a+\Lambda E_a{}^\mu)=&\; \tau_a{}^\mu -12\psibar(\lD_a\gamma^\mu-\gamma^\mu D_a)\Pi(\beta,\alpha)\psi \nonumber\\
 &-8i \epsilon_a{}^{bc\mu} \psibar \left(\lD_b \gamma_c+\gamma_c D_b\right)\Pi(-\beta,\alpha)\psi \nonumber\\
 &+4i \left[(T_{abc}+2T_{bac})\epsilon^{bcd\mu}-T^d{}_{bc}\epsilon_a{}^{bc\mu}\right]\psibar\gamma_d \Pi(-\beta,\alpha)\psi \nonumber\\
 &-2i R_{bcad}\epsilon^{bcd\mu} \phi_5+(R E_a{}^\mu+2G^\mu{}_a) \phi \nonumber\\
 &-12E_a{}^\mu (\phi^2+\phi_5^2)\,,\label{einsteineq}
\end{align}
where $\Pi(\beta,\alpha)=\beta \Pi_+ + \alpha \Pi_-$, $\slashed{R}=\half R^{ab}\Sigma_{ab}$, and $\tau_a{}^\mu$ is the energy momentum produced by the gauge fields $A$ and $A^I$, defined by $\delta L_\text{gauge} =\delta e^a_\mu \tau_a{}^\mu$.

These field equations can also be obtained combining the variations of the original action with respect to  $e^a$, $f^a$, $g^a$, $\psi$ and $\psibar$, before choosing the gravity sector \eqref{fgfixing}, as
\begin{align}
  \left(\frac{\delta \calL}{\delta e^a}+\rho \frac{\delta \calL}{\delta f^a}+\sigma \frac{\delta \calL}{\delta g^a} \right)-\half  i_{E_a}\left(\psibar \frac{\delta \calL}{\delta \psibar}+\frac{\delta \calL}{\delta \psi} \psi \right) =0\,,
\end{align}
where $i_X(\omega)$ is the contraction inner product of forms. Explicitly, this reads 
\begin{align} \label{C2} \nonumber
0 = &\; *\tau^a + \epsilon_{abcd}(\rho f^b - \sigma g^b)(R^{cd}+f^c f^d-g^c g^d) \\ \nonumber
 &+ 2i (\psibar \se)[\overleftarrow{D} \gamma_5 (\rho \gamma^a+\sigma \tilde{\gamma}^a)\se + \gamma_5 (\rho\gamma^a+\sigma \tilde{\gamma}^a) D](\se \psi) \\ \nonumber
 &+ 2i (\rho g^a + \sigma f^a)\psibar \se \se \psi\\ \nonumber
 &+ 2i \psibar (\lD \gf  [\gamma^a \Omega^- \se + \se \Omega^- \gamma^a]  + [\gamma^a \Omega^- \se + \se \Omega^- \gamma^a] D)\psi \\ \nonumber
 & + 2i f\cdot g\psibar [\gamma^a \se + \se \gamma^a] \psi + -4i \psibar [\gamma^a \gamma_5 \slashed{R} \se + \se \gamma_5 \slashed{R} \gamma^a] \psi \\ 
 & + 4i\psibar\gf(\sT \Omega^- \gamma^a -\gamma^a \Omega^- \sT)\psi + \frac{\ell^2}{3}\epsilon^a{}_{bcd}e^b e^c e^d (\phi' {}^2+\phi'_5 {}^2)\;,
\end{align} 
where $*\tau^a$ is the three-form dual of $\tau^a_\mu$. It can be readily seen that the dual of the above equation reduces to \eqref{einsteineq} in the gravity sector \eqref{fgfixing}, with the choice
$\alpha= (\rho +\sigma)/2$, $\beta=(\rho -\sigma)/2$. Note that equation \eqref{C2} transforms homogeneously under $e^a \to \kappa e^a$, $\psi \to \kappa^{-1} \psi$, a consequence of the Weyl invariance introduced by the definition \eqref{Afermion}, which is broken in the gravity sector \eqref{fgfixing}. Then, by taking the dual of this three form one obtains \eqref{einsteineq}.

The field equation for the matter field is given by the the Dirac equation,
\begin{align}
0 &\; = \; \gamma^a D_a \psi - \left[\half T^b{}_{ba} \gamma^a +\frac{i}{6} \epsilon^{abcd}T_{bcd} \gf\gamma_a\right] \psi \nonumber \\
& +\frac{1}{6\sqrt{2}\MP} \left[ \epsilon^{abcd}R_{abcd}\, i\gf + 2 R \right]\psi - \frac{1}{3 \MP^2} \left[\psibar\psi +(\psibar\gf\psi) \gf \right]\psi\,. \label{eqpsi3}
\end{align}

Here, the torsion coupling to the vector current has been restored after integration by parts. Suppressed by the factor $\MP^{-1}$, the non minimal coupling to the background curvature provides an effective mass term to the fermion. The NJL term, also suppressed by $\MP^{-2}$, can break chiral symmetry by a nonzero vacuum expectation value of the fermion condensate \cite{NJL}.


\end{document}